\documentclass[twocolumn,showpacs,prl]{revtex4}

\usepackage{graphicx}
\usepackage{amsmath}
\usepackage{amsfonts}
\usepackage{amssymb}
\usepackage{color}

\begin{document}

\newcommand{\bb}{_\mathrm{b}}
\newcommand{\pp}{_\mathrm{p}}
\newcommand{\sss}{_\mathrm{s}}
\newcommand{\rudd}{^\mathrm{RUDD}}
\newcommand{\udd}{^\mathrm{UDD}}

\title{High Order Coherent Control Sequences of Finite-Width Pulses}

\author{S. Pasini}
%\email{pasini@fkt.physik.tu-dortmund.de}
\affiliation{Lehrstuhl f\"{u}r Theoretische Physik I, 
Technische Universit\"{a}t Dortmund,
 Otto-Hahn Stra\ss{}e 4, 44221 Dortmund, Germany}

\author{P. Karbach}
\affiliation{Lehrstuhl f\"{u}r Theoretische Physik I, 
Technische Universit\"{a}t Dortmund,
 Otto-Hahn Stra\ss{}e 4, 44221 Dortmund, Germany}

\author{G. S. Uhrig}
\email{goetz.uhrig@tu-dortmund.de}
\affiliation{Lehrstuhl f\"{u}r Theoretische Physik I, 
Technische Universit\"{a}t Dortmund,
 Otto-Hahn Stra\ss{}e 4, 44221 Dortmund, Germany}

\date{\rm\today}

\begin{abstract}
The performance of sequences of designed pulses of finite length $\tau$
is analyzed for a bath of spins and it is compared with that of sequences 
of ideal, instantaneous pulses. The degree of the design of the pulse strongly 
affects the performance of the sequences. Non-equidistant, adapted sequences of pulses,
which equal instantaneous ones up to $\mathcal{O}(\tau^3)$, outperform equidistant 
or concatenated sequences. Moreover, they do so at low energy cost which
grows only logarithmically with the number of pulses, in contrast
to standard pulses with linear growth.
\end{abstract}


%% PACSes: to be redone - see also NMR
% 03.67.Pp Quantum error correction and other methods for protection against 
% decoherence (see also 03.65.Yz Decoherence; open systems; quantum statistical
%  methods; for decoherence in Bose-Einstein condensates, see 03.75.Gg) 
% 82.56.Jn Pulse sequences in NMR
% 03.67.Lx Quantum computation
% 76.60.Lz Spin echoes
%
% 03.65.Yz Decoherence; open systems; quantum statistical methods 
% (see also 03.67.Pp in quantum information; for decoherence in Bose-Einstein 
% condensates, see 03.75.Gg) 
% 03.65.Vf Phases: geometric; dynamic or topological  
% 82.56.Dj High resolution NMR
% 33.25.k Nuclear resonance and relaxation: Atomic and Molecular Physics

\maketitle

%\paragraph{Introduction.} 
The rapid evolution of the field of quantum science and quantum information demands 
robust quantum control techniques in the presence of environmental noise. 
To dynamically generate systems essentially free from decoherence has now become a 
focus of the research of quantum control. This suppression of decoherence is an
important requisite in quantum information processing 
\cite{niels00}, for example for the realization of a quantum computer, in nuclear
magnetic resonance (NMR), for high accuracy measurements \cite{haebe76} or in
 magnetic resonance imaging (MRI) \cite{jenis09}, to mention only a few. 

In this work we focus on quantum control by short pulses of finite length.
It is beyond our scope to discuss continuous quantum control, see for instance Ref.\
\onlinecite{gordo08}.
It is on a discovery in NMR, the Hahn spin echo \cite{hahn50}, that the 
pulsed-control methods are based. The original technique makes use of an electromagnetic 
pulse in order to rotate the spin and to refocus it along a desired direction. 
Dynamical decoupling (DD) \cite{viola98,ban98} 
iterates the single pulse in a sequence of pulses such that 
the coupling between the spin and its environment is averaged to zero. Among the 
\textquotedblleft open-loop\textquotedblright {} pulse-control techniques, the dynamical decoupling  is
 one of the most promising protocols for prolonging the coherence time of a 
spin (qubit) coupled to an environment. No detailed, quantitative knowledge of the decohering 
environment is required.

The sequences come in a large variety. We distinguish
equidistant and non-equidistant sequences. In the first category we recall the iterated 
Carr-Purcell-Meiboom-Gill (CPMG) sequence \cite{carr54,meibo58}, where  
the pulses are regularly separated (apart from the very first and the very 
last one). 
To the second category belong for instance the universal Uhrig DD (UDD) sequence 
\cite{uhrig07,lee08a,yang08}, the Locally Optimized Dynamical Decoupling (LODD) \cite{bierc09a}, 
the Optimized Noise Filtration by Dynamical Decoupling (OFDD) \cite{uys09} and the
Bandwidth-Adapted Dynamical Decoupling (BADD) \cite{khodj11a} for
 pure dephasing models and  the concatenated DD (CDD) \cite{khodj05} or UDD 
(CUDD) \cite{uhrig09b} or the quadratic UDD (QDD) \cite{west10} for 
models with dephasing and relaxation.

The design of the DD schemes relies originally on the assumption that the 
pulses are arbitrarily strong and instantaneous though the
effects of pulses of finite length were known to matter 
\cite{skinn03,viola03,sengu05,khodj07,pryad08a}.
But the pulses used in laboratories always have a bounded, finite amplitude
so that they have a finite duration $\tau$.
Even if sequences like CPMG and UDD have already been 
implemented in experiments with very good results \cite{bierc09a,du09,jenis09}, 
the fact that pulses have a finite duration appears often as a nuisance
deteriorating the suppression of decoherence, see for instance
Refs.\ \onlinecite{viola03} and \onlinecite{khodj07}.

It is of  great practical relevance to which extent the length of a pulse 
affects the performance of a sequence such as UDD or CPMG of given duration $T$. 
How should one choose the location, the duration (or the amplitude), and the 
shape \cite{sengu05,pryad08a,pasin09a} of the bounded pulse in order to
minimize the errors due to its finite duration if it replaces the ideal, 
instantaneous pulses in a certain sequence? 

Here we report for the first time numerical evidence of how sequences of 
realistic pulses of finite width must be designed in order to achieve the 
same perturbative suppression of dephasing as the corresponding ideal sequence. 
We compare various known sequences \cite{viola03,khodj09a, uhrig10a}
and  numerically analyze their  performance for a spin coupled to a bath of spins. 
To obtain an experimentally relevant comparison all pulses are designed in such a way 
that the  largest amplitude appearing in each sequence is the same \cite{enrg}.

\begin{figure}[ht]
    \begin{center}
    \includegraphics[width=0.35\columnwidth,clip]{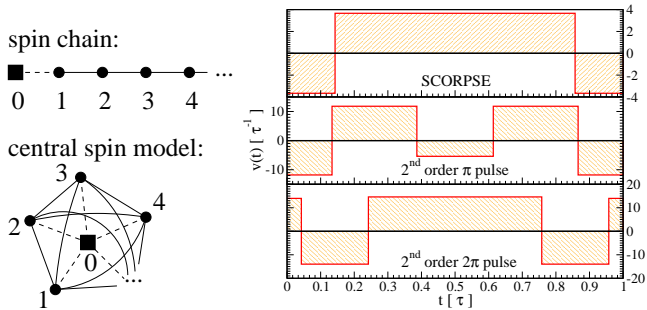}	
    \includegraphics[width=0.62\columnwidth,clip]{fig1b.eps}
    \end{center}
    \caption{Left: Spin bath models under study with a qubit (square)
    coupled (dashed lines) to $M$ bath spins (dots) interacting among themselves (solid lines).
    	Right: Upper panel: 1$^\text{st}$ order SCORPSE $\pi$ pulse \cite{cummi03};
      middle panel: 2$^\text{nd}$ order $\pi$ pulse; lower panel: 2$^\text{nd}$ order $2\pi$ pulse.
      Both 2$^\text{nd}$ order pulses are found by solving the 
    conditions derived in Ref.\ \onlinecite{pasin09a}. Amplitudes and
     switching instants are available upon request.
      \label{fig:pulse_shapes}
    }
\end{figure}

\paragraph{The Model.} We consider the pure dephasing Hamiltonian  
$H=1_q\otimes B_0+\sigma_z\otimes B_z$ that determines the free evolution of 
the system between two consecutive pulses by $U_{\mathrm{free}}(t)=\exp\{-itH\}$. 
The operators $B_0$ and $B_z$ act on the bath only, while the identity 
$1_q$ and the Pauli matrix $\sigma_z$ act on the qubit represented by
a spin $1/2$. For simplicity we identify henceforth $1_q\otimes B_0$
and $B_0$. The bath consists of $M$ spins with $i\in\{1,\ldots,M\}$
\begin{equation}
\label{eq:Hamiltonian} 
H = \omega_\mathrm{b}B_0 + 
\sigma^{(0)}_z\sum_{i=1}^{N_s}\lambda_i\ \sigma_z^{(i)}.
\end{equation} 
No drift term $\propto \sigma_z$ of the qubit 
is included because we work in the rotating reference frame.
Explicitly we analyze two cases, see also Fig.\ \ref{fig:pulse_shapes}: 
(i) A spin chain with $B_0=\sum_{i=1}^{M} \vec{\sigma}^{(i)}
\cdot\vec{\sigma}^{(i+1)}$, $\lambda_i\equiv \lambda$ and $N_s=1$.
(ii) A central spin model \cite{schli03,bortz07a,bortz07b,witze08,lee08a,witze10} 
characterized by a dipolar coupling \cite{haebe76} 
$B_0=\sum_{j<i=1}^{M}\left(3\sigma_z^{(i)} \sigma_z^{(j)}- \vec{\sigma}^{(i)}
\cdot\vec{\sigma}^{(j)}\right)$ with $\lambda_i (M-1)=\lambda(2i-M-1)$ and $N_s=M$.
The rapidity of the dynamics of the bath is given by 
$\omega_\mathrm{b}:=\alpha\lambda$ with $\alpha$ a dimensionless constant. 

The control Hamiltonian is given by $H_\mathrm{c}(t)=\sigma_x^0 v(t)$. 
We consider  piecewise constant pulses shown in Fig.\ \ref{fig:pulse_shapes}.
During each pulse of total length $\tau^{(i)}$ the qubit evolves under 
the simultaneous  action of the system and of the control Hamiltonian 
$U_{\mathrm{p}}={\cal T}
\exp\{-i \int_{t_0}^{t_0+\tau^{(i)}}(H+H_{\mathrm c}(t))dt\}$
where ${\cal T}$ stands for standard time ordering.
The evolution operator of the total sequence from $t=0$ to $t=T$ 
is denoted by $\widehat{R}$.

\begin{figure}[ht]
    \begin{center}
    \includegraphics[width=0.9\columnwidth,clip]{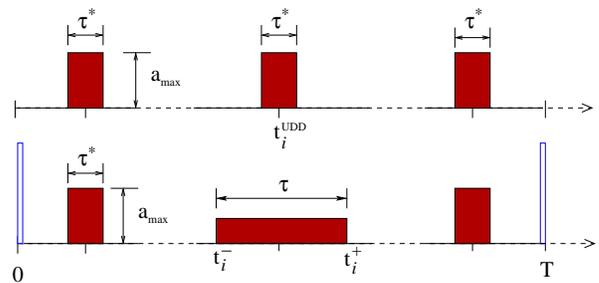}
    \end{center}
    \caption{(Color online) Upper panel:  Sequences of type (i) 
    ($\tau_j$CPMG, $\tau_j$CDD or $\tau_j$UDD) are sketched.
    	Lower panel: Sequences of type (ii) are shown ($\tau_j$RUDD).
    	 Only the maximum amplitude and the pulse duration are shown, 
    	 but no details of the pulse shapes. The $\pi$-pulses are depicted
    	 by filled (red) blocks, the initial and final $2\pi$-pulses 
    	 	by open (blue) blocks. The first $\pi$ 
      pulse of $\tau_j$RUDD and all pulses of type (i) sequences have the same 
      amplitude $a_{\mathrm{max}}$ to ensure experimentally relevant
      comparability. The instants 
      $t_i$ are given in Eqs.\ \ref{eq:inst_cpmg_udd} and \ref{eq:inst_cdd}; the 
      start and  end points $t_i^\pm$  in Eq.\ \eqref{eq:t_plus_minus}. 
      \label{fig1}
    }
\end{figure}

\paragraph{The Sequences.} Two types of sequences are studied, see also
Fig.\ \ref{fig1}:
(i) The durations $\tau^{(i)}=\tau^*$ of the pulses is constant throughout the
sequence and it is kept constant on variation of $T$. 
These sequences are denoted by $\tau_j$CPMG, $\tau_j$CDD and $\tau_j$UDD,
because they reproduce the ideal CPMG, CDD and UDD sequences for $\tau\rightarrow 0$.
The subscript $j$ stands for properties of the pulses as explained below.
(ii) The durations  $\tau^{(i)}$ are varied along the sequence,
 i.e., they depend on $i$. But they shall not depend on  $T$
other than that the sum of all pulse durations cannot exceed $T$, i.e.,
$T\ge T_\text{p}:=\sum_i \tau^{(i)}$. The corresponding sequences are
 denoted by $\tau_j$RUDD.

The  sequences of type (i) are made of $N$ $\pi$ pulses whose  
width is $\tau^*$. The center of the $i$-th pulse is given by
\begin{subequations}
\label{eq:inst_cpmg_udd}
\begin{equation}
 \label{eq:inst_cpmg}
t_i^\text{CPMG}:=T\ (2i-1)/(2N),
\end{equation}  
\begin{equation}
 \label{eq:inst_udd}
t_i^\text{UDD}:=T\ \sin^2\left(\pi\, i/(2(N+1))\right),
\end{equation} 
\end{subequations}
for the $\tau_j$CPMG and $\tau_j$UDD \cite{uhrig07} sequence, respectively.
We use the simplified version of CDD designed only for pure dephasing.
The CDD sequence of level $k$ is defined by the recursion
\begin{subequations}
\label{eq:inst_cdd}
\begin{equation}
\label{eq:cdd-even}
\text{CDD}_{k+1}(T)=\text{CDD}_{k}\left({T/2}\right)\circ 
\Pi_{\pi} \circ \text{CDD}_{k}\left({T/2}\right),\\
\end{equation}
\begin{equation}
\label{eq:cdd-odd}
\text{CDD}_{k+1}(T)=\text{CDD}_{k}\left({T/2}\right)\circ\text{CDD}_{k}\left({T/2}\right),
\end{equation}
\end{subequations}
where \eqref{eq:cdd-even} holds for $k$ even and \eqref{eq:cdd-odd}
for $k$ odd; $\circ$ stands for concatenation and $\Pi_{\varphi}$ 
 for the operator of a pulse of angle $\varphi$. 
The zero-level CDD$_{0}(T)$ is free evolution without pulses.

The subscript $j$ in $\tau_{j}$ refers to the order of the pulses, i.e.,
its time evolution operator fulfills 
$U_{\mathrm{p}}=\exp\{-i \tau B_0\}\Pi_{\varphi} +\mathcal{O}(\tau^{j+1})$.
We restrict our study here to explicit pulses with $j=0, 1, 2$, see Fig.\
\ref{fig:pulse_shapes}, which fulfill
the conditions derived in Ref.\ \onlinecite{pasin09a}.
A recursion for general $j$ is given in Ref.\ \onlinecite{khodj10}.
The 0$^\text{th}$ order pulse is simply rectangular; the other pulses used 
are depicted in Fig.\ \ref{fig:pulse_shapes}.

The sequences of type (ii) are similar to the $\tau_j$UDD sequences in that they
are based on pulses of order $j$. The crucial difference is 
that pulses are not constant in length. They are defined according
to our previous work \cite{uhrig10a} by a
start instant $t_i^-$ and a stop instant $t_i^+$
given by 
\begin{equation}
\label{eq:t_plus_minus}
t^{\pm}_{i}:=
T \sin^2\left(\frac{\pi\  i}{2(N+1)}
\pm\frac{\theta_\mathrm{p}(T)}{2}\right).
\end{equation}
The above relation results naturally
from the requirement that the effective switching
function of the sequence expressed in $\theta\in [0,\pi ]$
according to $t=T\sin^2(\theta/2)$ is antiperiodic \cite{uhrig10a}.
This antiperiodicity ensures that the total sequence
suppresses the decohering terms $\propto \sigma_z$ in the time evolution \cite{yang08}.
The duration of the pulses in time $\tau^{(i)}= t^+_{i}-t^-_{i}$ yielding
\begin{equation}
\label{eq:tau_p}
\tau^{(i)}= T\sin\left(\pi\  i/(N+1)\right)\sin(\theta_\mathrm{p})
\end{equation}
is determined by the parameter $\theta_\mathrm{p}(T)$. It
acquires a dependence on $T$ if we require $\tau^*:=\tau^{(1)}$
to be constant upon varying $T$.
Note that $\theta_\mathrm{p}=\pi/(2(N+1))$ refers to back-to-back pulses
without any free evolution between them, see below.

The antiperiodicity of the switching function 
is the basis for the suppression of dephasing in high order
\cite{yang08,uhrig10b}. In order to guarantee this antiperiodicity,
it is required to insert an initial and a final pulse which
represent the identity 
$U_{\mathrm{p}}=\exp\{-i \tau B_0\} +\mathcal{O}(\tau^{j+1})$.
For instance, it may be a zero $\pi$ or a $2\pi$ pulse \cite{uhrig10a}.
The initial pulse starts at $t_0^-=0$ and stops at 
$t_0^+=T \sin^2\left(\theta_\mathrm{p}/2\right)$ while  the final one
starts at $t_{N+1}^-=T \sin^2\left[(\pi-\theta_\mathrm{p})/2\right]$ and stops 
at $t_{N+1}^+=T$. These pulses are indicated by open boxes in Fig.\ \ref{fig1}.

In the sequel, we compare the various sequences always with the
same $\tau^*$ because the shortest accessible pulse duration of a $\pi$ pulse, corresponding
to the maximum amplitude, represents a crucial experimental constraint \cite{enrg,khodj11a}.
Only the very short boundary $2\pi$ pulses in the $\tau_j$RUDD are treated separately.
But their importance is assessed by considering $\tau_j$RUDD with and without 
the boundary $2\pi$ pulses.
We stress that due to the variable duration of the pulses according
to (\ref{eq:t_plus_minus},\ref{eq:tau_p}) in the RUDD sequence
most of the pulses are much longer than $\tau^*$.

\paragraph{The Partial Frobenius ($\Delta_\mathrm{pF}$) Distance} 
defines the distance between the 
ideal evolution of the initial state of the qubit due to the
pulses and its evolution including the interaction with the bath and the application
of the sequence \cite{lidar08}.  
For each axis of rotation $\gamma=\{x,y,z\}$ we define a difference
of density matrices of the qubit by
$ \rho_\mathrm{q}^{(\gamma)}:=\mathrm{tr_B}\left[\rho_\mathrm{id}^{(\gamma)}
-\rho_\mathrm{qB}^{(\gamma)}\right]$,
where $\rho_\mathrm{qB}^{(\gamma)}:=\widehat{R}\rho_0^{(\gamma)}\widehat{R}^{\dagger}$. 
The partial trace over the bath is denoted by $\mathrm{tr_B}$. 
Given a factorized initial state 
$\rho_0^{(\gamma)}:=|\gamma\rangle\langle\gamma| \otimes 1_\mathrm{B}$ the density matrix $\rho_\mathrm{id}^{(\gamma)}:=
\sigma_x^N\rho_0^{(\gamma)}\sigma_x^N$ is the ideally evolved $\rho_0$ 
subject only  to ideal pulses without any bath interaction.
The distance $\Delta_\mathrm{pF}$ measures  the difference between the real evolution and
the ideal one reading
\begin{equation}
 \label{eq:tr_norm} 
\Delta_\mathrm{pF}^2:={\frac{1}{3}\sum_{\gamma=x,y,z}\mathrm{tr_q} 
\left[\rho_\mathrm{q}^{(\gamma)}\right]^2}.
\end{equation}

\begin{figure}[ht]
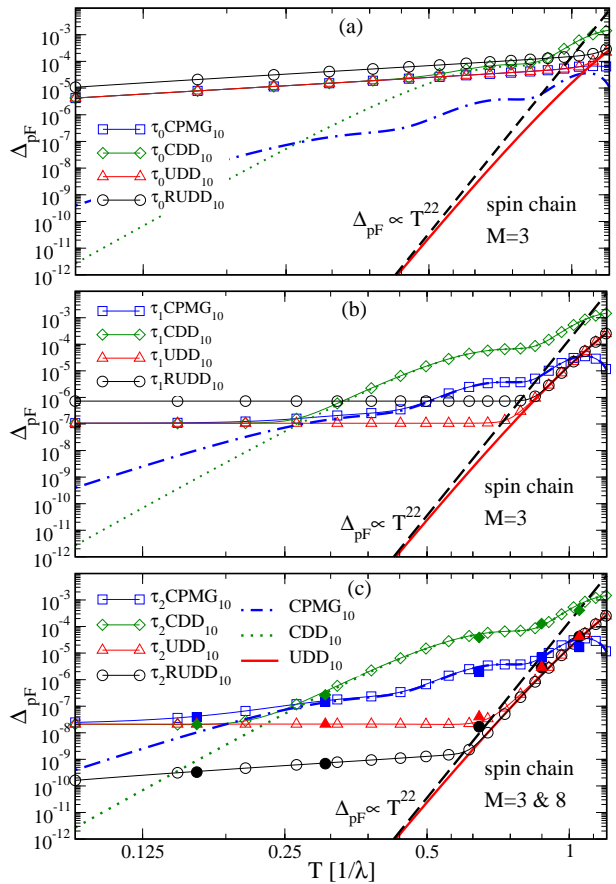

    \includegraphics[width=0.93\columnwidth,clip]{fig3a.eps}
    \includegraphics[width=0.93\columnwidth,clip]{fig3b.eps}
    \includegraphics[width=0.93\columnwidth,clip]{fig3c.eps}
    \caption{(Color online) Distance $\Delta_\mathrm{pF}$  vs.\ the duration $T$ of sequences
    	of pulses with zero or finite width for $\alpha=10$ for spin chains. 
    		All open symbols refer to $M=3$ bath spins; the filled symbols in panel (c) to $M=8$ bath spins.
       The finite-width pulses have minimum width $\tau^*=1.086\cdot 10^{-3}/\lambda$ and $T\ge 0.09/\lambda$;
       Panel (a): rectangular 0$^\mathrm{th}$ order pulses;
        Panel (b): SCORPSE 1$^\mathrm{st}$ order pulses \cite{cummi03};        
        Panel (c): 2$^\mathrm{nd}$ order pulses shown in Fig.\ \ref{fig:pulse_shapes}.
        	To highlight power-law behavior the dashed lines are included: The UDD curve scales as 
        		$T^{2(N+1)}$.
    \label{fig3}
    }
\end{figure}

\begin{figure}[ht]
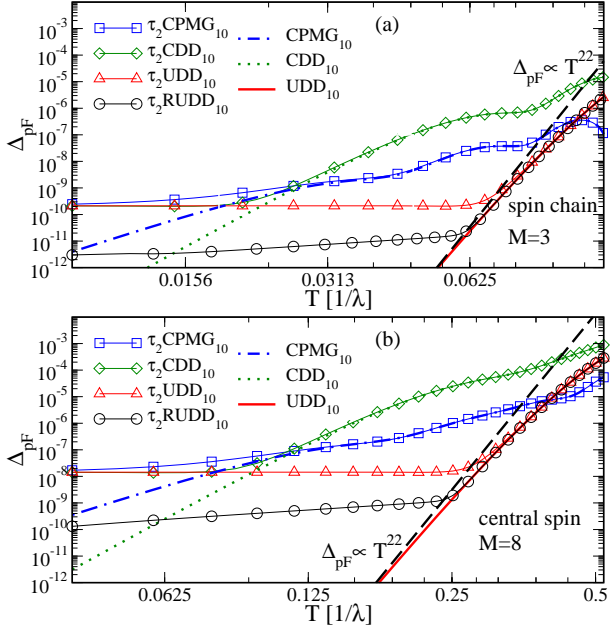

    \includegraphics[width=0.93\columnwidth,clip]{fig4a.eps}
    \includegraphics[width=0.93\columnwidth,clip]{fig4b.eps}
    \caption{(Color online) Same as in Fig.\ \ref{fig3}(c). Panel (a): spin chain with
    $M=3$, $\alpha=100$ and $\tau^*=1.086\cdot 10^{-4}/\lambda$, $T\ge 0.009/\lambda$.
    Panel (b): central spin model
    	with $M=8$, $\alpha=10$ and $\tau^*=0.0004828/\lambda$, $T\ge 0.04/\lambda$. 
    	The result is identical to Fig.\ \ref{fig3}(c) except for a shift by
    	the factor $\kappa\approx 2.3$ in $T$.
    \label{fig4}
    }
\end{figure}

\begin{figure}[ht]
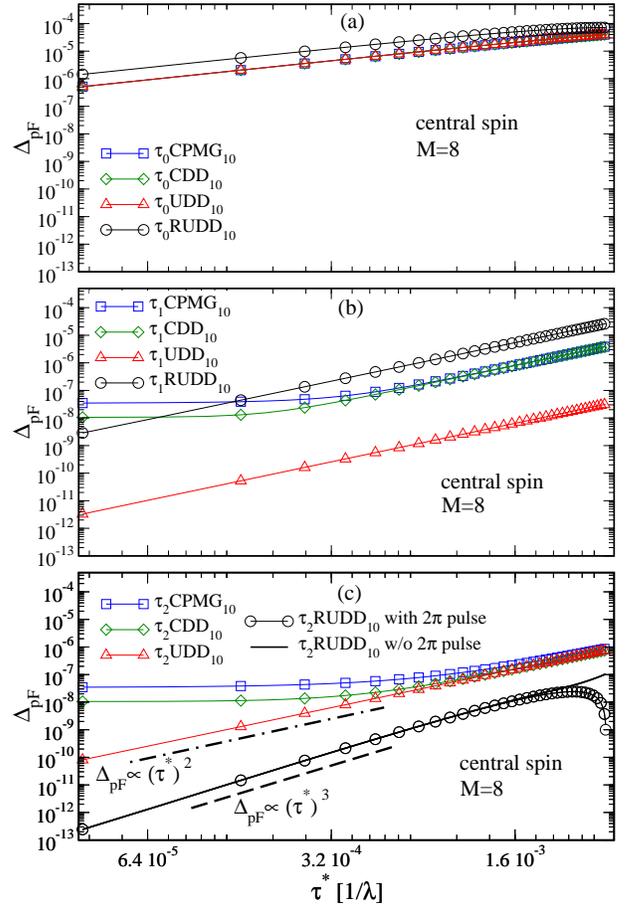

    \includegraphics[width=0.93\columnwidth,clip]{fig5a.eps}
    \includegraphics[width=0.93\columnwidth,clip]{fig5b.eps}
    \includegraphics[width=0.93\columnwidth,clip]{fig5c.eps}
    \caption{(Color online) Distance $\Delta_\textrm{pF}$ vs.\ 
    	the shortest pulse duration $\tau^*$ for
       various sequences at $\alpha=10$ with $T=0.09/\lambda$;
       the panels correspond to pulses of finite width of different order as in Fig.\ \ref{fig3}.
    \label{fig5}
    }
\end{figure}

\paragraph{Numerical Simulation.} 
We compute the performance of sequences of pulses of finite duration
for the systems in  (\ref{eq:Hamiltonian}) shown in Fig.\ \ref{fig:pulse_shapes}. 
We choose the minimum duration $\tau^*<\min_i\{\tau^{(i)}\}$ and a minimum value 
of $T$ such that $T\ge \sum_i \tau^{(i)}$, see captions for values. 
Sequences with $N=10$ pulses are considered because this number
 allows us to consider the CDD sequence as well; it corresponds to the concatenation level $k=4$,
 cf.\  Eq.\ \eqref{eq:inst_cdd}. The results are shown in Figs.\ \ref{fig3} and \ref{fig4}(a) for the spin chain model
 and in Figs.\ \ref{fig4}(b) and Fig.\ \ref{fig5} for the central spin model.

First, we consider the influence of the topology and the size of the spin bath. 
In Fig.\ \ref{fig3}(c) data for the spin chain is shown for $M=3$ (open symbols) and data
for $M=8$ (filled symbols) fits in perfectly. This indicates that the size effect is very small in
the regime of interest. The  topology of the spin bath
has a certain impact, but only on the quantitative level, not on the qualitative one
as can be seen comparing Fig.\ \ref{fig3}(c) with Fig.\ \ref{fig4}(b). The results for the 
central spin model with $M=8$ bath spins are qualitatively identical to the ones for the spin chain
except for a heuristic factor $\kappa\approx 2.3$ in $T$. The latter can easily be understood in the sense
of an effectively stronger coupling between qubit and bath for the central spin model than
for the spin chain for the same value $\lambda$ because there are \emph{more} couplings 
$\lambda_i\propto\lambda$ between qubit and bath spins.

Second, we study the influence of the sequences on the performance. Thus we
consider long sequence durations $T$. In this regime the pulse errors are unimportant and
pulse shaping plays only a minor role. This fact is perfectly understandable
because for given $\tau^*$ the limit $T\to\infty$ implies that $\tau^*/T$ vanishes.
In the formalism of filter functions \cite{uhrig07,uhrig08,cywin08,bierc09a,bierc11a}
this can easily be seen. The signal $s(T)=\exp(-2\chi(T))$ is determined by the frequency integral
\begin{equation}
 \label{eq:chi}
 \chi(T):=\int_0^{\infty}\frac{S(\omega)}{\omega^2}F(\omega T)d\omega
\end{equation}
where $F(\omega T)$ is the filter function. For pulses of duration $\tau^{(j)}$
centered at instants $\delta_j T$ it is given by
\begin{equation}
\label{eq:F_deltaPi}
 F(z)=\Big|1+(-1)^{N+1}e^{-iz}+2\sum_{j=1}^N e^{i z \delta_j}
 \cos\big(\frac{z \tau^{(j)}}{2T}\big)\Big|^2,
\end{equation}
where we use $z:=\omega T$ for brevity. This equation is valid if the coupling
between qubit and bath is effectively zero during the pulse. For artifical
noise this can be realized experimentally \cite{bierc09a} while for generic
systems the pulse design has to approximate this situation \cite{pasin09a,uhrig10a}.
Clearly, for larger and larger $T$ the influence of the finite pulse durations
$\tau^{(j)}$ decreases more and more.

% parity

The scaling of $\Delta_\mathrm{pF}$ with $T$ for UDD with ideal pulses
is also remarkable. For UDD, $d\lessapprox |b_{+-}|+ |b_{--}|$ 
where $b_{+-}$ and $b_{--}$ depend on $B_0$, $B_z$ 
and on the initial density matrix $\rho_0$ \cite{uhrig10b}.
In particular, $b_{--}\varpropto T^{2(N+1)}$ and its prefactor is even in $B_z$ while 
$b_{+-}\varpropto T^{(N+1)}$ with a prefactor odd in $B_z$. If both are present
one has the generic result $d=\mathcal{O}(T^{N+1})$. But if the Hamiltonian is symmetric
under global spin flip $\sigma_z \leftrightarrow -\sigma_z$, realized, e.g.,
by a $\pi$ rotation about total $\sigma_x$, it follows that $b_{+-}=-b_{+-}=0$
due to its oddness in $B_z$ such that we obtain
$d=\mathcal{O}(T^{2(N+1)})$ which is better than generically expected. 
Hence on the one hand, the generic behavior of
dynamic decoupling can only be seen for systems without symmetry.
On the other hand, we stick here to the Hamiltonian \eqref{eq:Hamiltonian}
because it is of the kind occurring mostly in experiment \cite{haebe76,schli03,levit05}.

Fig.\ \ref{fig3}(c) with $\alpha=10$ and Fig.\ \ref{fig4}(a) with $\alpha=100$ 
differ in the rapidity of the bath dynamics which is faster for larger $\alpha$. 
Clearly, the decoherence sets in earlier if the bath is faster because the
switching by the pulses is relatively slower.
This is no contradiction to the basic idea of motional narrowing stating that a
very fast bath  implies longer coherence times because the fast bath
dynamics reduces its influence on the qubit due to averaging. 
But previous results, e.g., Fig.\ 3 in Ref.\ \onlinecite{west10}, 
show that for this effect to take place $\alpha$ should exceed $10^6$.

We do not consider data for smaller $\alpha\lessapprox 1$ here because
it is the our present scope to show how the detrimental effect
of finite pulse duration can be compensated.
But a previous study on single pulses, see Fig.\ 7 in  Ref.\ \onlinecite{karba08},
revealed that effects of the finite duration of 
the pulses become noticeable only for $\alpha>1$.

Third, we consider the large regime of shorter durations $T$
where $\Delta_\mathrm{pF}$ is dominated by the properties of the pulses. 
Naturally, this effect is most prominent
for the uncorrected rectangular pulses of $0^\mathrm{th}$ order. In Fig.\ \ref{fig3}(a)
the distance $d$ is significantly larger for pulses of finite width (symbols) than 
for the ideal ones (lines). The RUDD sequence performs worse than the other sequences.
This is not surprising since it is based on the assumption that
the pulse is designed such that there is none or no significant coupling
between qubit and bath during the pulse. A rectangular pulse realizes
this assumption only in order $\tau^*$.

Hence it is clear that the level for $\Delta_\mathrm{pF}$ which can be reached for small
values of $T$ is lower for the $1^\mathrm{st}$ order pulses (panel (b)) and even lower
for the $2^\mathrm{nd}$ order pulses (panel (c)). This fact illustrates nicely
that the optimization of pulses is indeed an important ingredient
in enhancing the performance of dynamic decoupling 
\cite{sengu05,pryad08a,pasin09a,khodj10}.

The key observation is that the $\tau_j$RUDD becomes
the best performing for $j=2$. For $j=0$ and $j=1$ the $\tau_j$UDD
sequence turned out to be more advantageous. We conclude that the pulses
need to be sufficiently well designed in order that the underlying
idea of the RUDD sequence \cite{uhrig10a} really pays. In Fig.\ \ref{fig3}(c)
the gain using RUDD instead of UDD is about two orders of magnitude.
Such improvements are to be expected in the regime where the performance
of the sequences is dominated by the pulse errors.

We emphasize that the fact that RUDD performs better than UDD or any other generic sequence
of pulses of constant duration is quite remarkable because most of the 
pulses in the RUDD sequence are much \emph{longer} than $\tau^*$.
The sum $T_\text{p}$ 
of the lengths of all $N$ pulses is $T_\text{p}=N\tau^*$ for a generic sequence  while it is 
\begin{subequations}
\begin{align}
T_\text{p} & =\tau^* {\cot(\pi/(2(N+1))}/{\sin(\pi/(N+1))}
\\
& \approx \tau^* {2(N+1)^2}/{\pi^2} \qquad \text{for} \ N \ \text{large}
\end{align}
\end{subequations}
for the RUDD sequence according to Eqs.\ (\ref{eq:t_plus_minus},\ref{eq:tau_p}).
One may prefer to consider the total energy necessary to realize the sequence \cite{gordo08}.
The energy required for a given pulse is  proportional to $1/\tau$. Hence
the total energy $E_\text{p}$ is given for the UDD sequence by $E_\text{p}=A N/\tau^*$
where $A$ is a constant depending on the shape of the pulse. Note the \textit{linear}
divergence in $N$. In contrast, for the RUDD sequence one obtains
\begin{subequations}
\begin{align}
E_\text{p} & =\frac{A\sin(\pi/(N+1))}{\tau^*} \sum_{j=1}^N \frac{1}{\sin(\pi j/(N+1))}
\\
& \approx ({2A}/{\tau^*}) \ln\left[{2(N+1)}/{\pi} \right]\qquad \text{for} \ N \ \text{large}
\end{align}
\label{eq:energy_rudd}
\end{subequations}
which diverges only \textit{logarithmically} in $N$. Thus, given a minimum pulse duration
$\tau^*$ it is much less costly in energy to reach long coherence times by
applying RUDD than by any generic sequence with pulses of constant $\tau^*$.

In view of the above observations, it remains to clarify why the RUDD works better 
than the other sequences, but only for higher order pulses. According to the
analytic foundation of RUDD \cite{uhrig10a}, its advantage over
other sequences with shaped pulses consists in the vanishing of mixed terms
in $T$ and $\tau^*$. For instance, an ideal UDD$_N$ scales generically like $T^{N+1}$
and the $\tau_j$UDD$_N$ of $N$ finite-width pulses certainly has errors scaling like
$T^{N+1}$ and $(\tau^*)^{j+1}$. But one cannot exclude the occurrence of 
terms such as $T\tau^*$, $T^2\tau^*$, or  $T(\tau^*)^2$. They result
from the interplay between the finite duration of the pulses and the sequence.
It is crucial that this is different for $\tau_j$RUDD$_N$. There the finite
duration is fully taken into account in the design of the sequence \cite{uhrig10a}.
Hence the errors of the $\tau_j$RUDD$_N$ are of the order $T^{N+1}$ and $(\tau^*)^{j+1}$;
the lowest mixed terms are $T^{N+1}\tau^*$ and $T(\tau^*)^{j+1}$.

The above argument lays the foundation why the RUDD outperforms other sequences.
To illustrate the argument we plot the dependence of $\Delta_\mathrm{pF}$ on $\tau^*$ for
various sequences of finite-width pulses in Fig.\ \ref{fig5}
 for the central spin model at $M=8$. Results for the spin chain model
(not shown) look very much the same except for a rescaling of $T$.
In panel (a) all sequences behave similarly; the dependence on $\tau^*$ is
linear, and the RUDD behaves worst. This fact is attributed
to the larger average length of the pulses. Note that in the regime
depicted the distance $\Delta_\textrm{pF}$ is still fully dominated by the pulse errors.

In panel (b) we can nicely see the crossover from the regime
where the pulse error dominates (straight lines corresponding
to $(\tau^*)^2$) to the saturation levels corresponding to the
errors of the ideal CPMG and CDD sequence. The errors of the ideal
UDD sequence is much lower so that its saturation level cannot
be seen. Still the $\tau_1$RUDD behaves worse than the $\tau_1$UDD.

In panel (c) we again see the crossover from pulse errors to 
sequence errors on $\tau^*\to 0$. Interestingly, the RUDD behaves
better than the UDD in that the pulse errors decrease expectedly faster 
$\Delta_\text{pF,RUDD}\propto (\tau^*)^3$ compared to 
$\Delta_\text{pF,UDD}\propto (\tau^*)^2$. We stress that the latter
scaling is no contradiction to the pulse being second order
because an error $T(\tau^*)^2$ is not excluded. Fig.\ \ref{fig5}(c)
establishes that such mixed terms indeed deteriorate the performance
of unadapted sequences of finite-width pulses.  This clarifies the behavior
of RUDD relative to other sequences.

For practical implementation, 
it is important to point out that the behavior of $\tau_2$RUDD for small
$\tau^*$ is independent of whether or not we include the very short
boundary $2\pi$ pulses, cf.\ solid line and circles in Fig.\ \ref{fig5}(c).
This is due to the shortness of these effective identity pulses.

Last but not least, we find another regime of low values of $\Delta_\mathrm{pF}$.
This is the regime where
the pulse lengths reach their maximum value because the pulses
touch one another. They are back to back. Quite unexpectedly, the
full RUDD including the boundary $2\pi$ pulses again permits to obtain
an extremely good suppression of decoherence. This regime is
very interesting because it requires only very low pulse amplitudes
and a small total energy for the coherent control, cf.\
Eq.\ \eqref{eq:energy_rudd}, due to the pulses of maximum length.
Further studies of this relevant regime are left to future research.

%%%

\paragraph{Conclusions.} 
The analysis of sequences of finite-width pulses allows us to draw the 
following conclusions. They are derived from the data for the models
studied, but we expect them to hold more generally.

First, the use of
higher order pulses generically implies a significant improvement. 
Such pulses are designed such that they suppress the coupling to the bath to a 
high order during their action  \cite{pasin09a}. 
Second, non-equidistant sequences such as UDD outperform or, 
in the worst case, perform the same as
 equidistant (CPMG) or concatenated (CDD) sequences. 

Third, in the regime, where the pulse errors dominate the suppression of decoherence
is further enhanced by varying the pulse durations along the sequence (RUDD) as suggested
on analytic grounds \cite{uhrig10a}. This enhancement takes only place for pulses of 
sufficient high order. We found that it is present for second order pulses.
This establishes RUDD as a promising concept and represents our
central result.

Fourth, an additional interesting asset of
the RUDD is that the total energy required for the coherent control by pulses
increases only logarithmically with the number of pulses -- in contrast to
all other sequences of unvaried pulses. Hence in particular long coherence times
can be realized at low energy price.

Fifth, surprisingly, we found an additional regime where the RUDD suppresses decoherence efficiently.
This is the regime where the pulses are (almost) back-to-back approaching continuous
modulation \cite{gordo08}. Because in this regime the pulses reach their maximum length
the required control energy is a minimum.
Further research is required to study this promising regime in detail.

%\bibliographystyle{apsrev}
%\bibliography{../../bibinput/liter10}

\begin{thebibliography}{44}
\expandafter\ifx\csname natexlab\endcsname\relax\def\natexlab#1{#1}\fi
\expandafter\ifx\csname bibnamefont\endcsname\relax
  \def\bibnamefont#1{#1}\fi
\expandafter\ifx\csname bibfnamefont\endcsname\relax
  \def\bibfnamefont#1{#1}\fi
\expandafter\ifx\csname citenamefont\endcsname\relax
  \def\citenamefont#1{#1}\fi
\expandafter\ifx\csname url\endcsname\relax
  \def\url#1{\texttt{#1}}\fi
\expandafter\ifx\csname urlprefix\endcsname\relax\def\urlprefix{URL }\fi
\providecommand{\bibinfo}[2]{#2}
\providecommand{\eprint}[2][]{\url{#2}}

\bibitem[{\citenamefont{Nielsen and Chuang}(2000)}]{niels00}
\bibinfo{author}{\bibfnamefont{M.~A.} \bibnamefont{Nielsen}} \bibnamefont{and}
  \bibinfo{author}{\bibfnamefont{I.~L.} \bibnamefont{Chuang}},
  \emph{\bibinfo{title}{Quantum Computation and Quantum Information}}
  (\bibinfo{publisher}{Cambridge University Press},
  \bibinfo{address}{Cambridge}, \bibinfo{year}{2000}).

\bibitem[{\citenamefont{Haeberlen}(1976)}]{haebe76}
\bibinfo{author}{\bibfnamefont{U.}~\bibnamefont{Haeberlen}},
  \emph{\bibinfo{title}{High Resolution NMR in Solids: Selective Averaging}}
  (\bibinfo{publisher}{Academic Press}, \bibinfo{address}{New York},
  \bibinfo{year}{1976}).

\bibitem[{\citenamefont{Jenista et~al.}(2009)\citenamefont{Jenista, Stokes,
  Branca, and Warren}}]{jenis09}
\bibinfo{author}{\bibfnamefont{E.~R.} \bibnamefont{Jenista}},
  \bibinfo{author}{\bibfnamefont{A.~M.} \bibnamefont{Stokes}},
  \bibinfo{author}{\bibfnamefont{R.~T.} \bibnamefont{Branca}},
  \bibnamefont{and} \bibinfo{author}{\bibfnamefont{W.~S.}
  \bibnamefont{Warren}}, \bibinfo{journal}{J. Chem. Phys.}
  \textbf{\bibinfo{volume}{131}}, \bibinfo{pages}{204510}
  (\bibinfo{year}{2009}).

\bibitem[{\citenamefont{Gordon et~al.}(2008)\citenamefont{Gordon, Kurizki, and
  Lidar}}]{gordo08}
\bibinfo{author}{\bibfnamefont{G.}~\bibnamefont{Gordon}},
  \bibinfo{author}{\bibfnamefont{G.}~\bibnamefont{Kurizki}}, \bibnamefont{and}
  \bibinfo{author}{\bibfnamefont{D.~A.} \bibnamefont{Lidar}},
  \bibinfo{journal}{Phys. Rev. Lett.} \textbf{\bibinfo{volume}{101}},
  \bibinfo{pages}{010403} (\bibinfo{year}{2008}).

\bibitem[{\citenamefont{Hahn}(1950)}]{hahn50}
\bibinfo{author}{\bibfnamefont{E.~L.} \bibnamefont{Hahn}},
  \bibinfo{journal}{Phys. Rev.} \textbf{\bibinfo{volume}{80}},
  \bibinfo{pages}{580} (\bibinfo{year}{1950}).

\bibitem[{\citenamefont{Viola and Lloyd}(1998)}]{viola98}
\bibinfo{author}{\bibfnamefont{L.}~\bibnamefont{Viola}} \bibnamefont{and}
  \bibinfo{author}{\bibfnamefont{S.}~\bibnamefont{Lloyd}},
  \bibinfo{journal}{Phys. Rev. A} \textbf{\bibinfo{volume}{58}},
  \bibinfo{pages}{2733} (\bibinfo{year}{1998}).

\bibitem[{\citenamefont{Ban}(1998)}]{ban98}
\bibinfo{author}{\bibfnamefont{M.}~\bibnamefont{Ban}}, \bibinfo{journal}{J.
  Mod. Opt.} \textbf{\bibinfo{volume}{45}}, \bibinfo{pages}{2315}
  (\bibinfo{year}{1998}).

\bibitem[{\citenamefont{Carr and Purcell}(1954)}]{carr54}
\bibinfo{author}{\bibfnamefont{H.~Y.} \bibnamefont{Carr}} \bibnamefont{and}
  \bibinfo{author}{\bibfnamefont{E.~M.} \bibnamefont{Purcell}},
  \bibinfo{journal}{Phys. Rev.} \textbf{\bibinfo{volume}{94}},
  \bibinfo{pages}{630} (\bibinfo{year}{1954}).

\bibitem[{\citenamefont{Meiboom and Gill}(1958)}]{meibo58}
\bibinfo{author}{\bibfnamefont{S.}~\bibnamefont{Meiboom}} \bibnamefont{and}
  \bibinfo{author}{\bibfnamefont{D.}~\bibnamefont{Gill}},
  \bibinfo{journal}{Rev. Sci. Inst.} \textbf{\bibinfo{volume}{29}},
  \bibinfo{pages}{688} (\bibinfo{year}{1958}).

\bibitem[{\citenamefont{Uhrig}(2007)}]{uhrig07}
\bibinfo{author}{\bibfnamefont{G.~S.} \bibnamefont{Uhrig}},
  \bibinfo{journal}{Phys. Rev. Lett.} \textbf{\bibinfo{volume}{98}},
  \bibinfo{pages}{100504} (\bibinfo{year}{2007});
  \bibinfo{journal}{Erratum:} \textbf{\bibinfo{volume}{106}},
  \bibinfo{pages}{129901} (\bibinfo{year}{2011}{\natexlab{a}}).

\bibitem[{\citenamefont{Lee et~al.}(2008)\citenamefont{Lee, Witzel, and
  \mbox{Das~Sarma}}}]{lee08a}
\bibinfo{author}{\bibfnamefont{B.}~\bibnamefont{Lee}},
  \bibinfo{author}{\bibfnamefont{W.~M.} \bibnamefont{Witzel}},
  \bibnamefont{and}
  \bibinfo{author}{\bibfnamefont{S.}~\bibnamefont{\mbox{Das~Sarma}}},
  \bibinfo{journal}{Phys. Rev. Lett.} \textbf{\bibinfo{volume}{100}},
  \bibinfo{pages}{160505} (\bibinfo{year}{2008}).

\bibitem[{\citenamefont{Yang and Liu}(2008)}]{yang08}
\bibinfo{author}{\bibfnamefont{W.}~\bibnamefont{Yang}} \bibnamefont{and}
  \bibinfo{author}{\bibfnamefont{R.-B.} \bibnamefont{Liu}},
  \bibinfo{journal}{Phys. Rev. Lett.} \textbf{\bibinfo{volume}{101}},
  \bibinfo{pages}{180403} (\bibinfo{year}{2008}).

\bibitem[{\citenamefont{Biercuk et~al.}(2009)\citenamefont{Biercuk, Uys,
  VanDevender, Shiga, Itano, and Bollinger}}]{bierc09a}
\bibinfo{author}{\bibfnamefont{M.~J.} \bibnamefont{Biercuk}},
  \bibinfo{author}{\bibfnamefont{H.}~\bibnamefont{Uys}},
  \bibinfo{author}{\bibfnamefont{A.~P.} \bibnamefont{VanDevender}},
  \bibinfo{author}{\bibfnamefont{N.}~\bibnamefont{Shiga}},
  \bibinfo{author}{\bibfnamefont{W.~M.} \bibnamefont{Itano}}, \bibnamefont{and}
  \bibinfo{author}{\bibfnamefont{J.~J.} \bibnamefont{Bollinger}},
  \bibinfo{journal}{Nature} \textbf{\bibinfo{volume}{458}},
  \bibinfo{pages}{996} (\bibinfo{year}{2009}).

\bibitem[{\citenamefont{Uys et~al.}(2009)\citenamefont{Uys, Biercuk, and
  Bollinger}}]{uys09}
\bibinfo{author}{\bibfnamefont{H.}~\bibnamefont{Uys}},
  \bibinfo{author}{\bibfnamefont{M.~J.} \bibnamefont{Biercuk}},
  \bibnamefont{and} \bibinfo{author}{\bibfnamefont{J.~J.}
  \bibnamefont{Bollinger}}, \bibinfo{journal}{Phys. Rev. Lett.}
  \textbf{\bibinfo{volume}{103}}, \bibinfo{pages}{040501}
  (\bibinfo{year}{2009}).

\bibitem[{\citenamefont{Khodjasteh et~al.}(2011)\citenamefont{Khodjasteh,
  Erd\'elyi, and Viola}}]{khodj11a}
\bibinfo{author}{\bibfnamefont{K.}~\bibnamefont{Khodjasteh}},
  \bibinfo{author}{\bibfnamefont{T.}~\bibnamefont{Erd\'elyi}},
  \bibnamefont{and} \bibinfo{author}{\bibfnamefont{L.}~\bibnamefont{Viola}},
  \bibinfo{journal}{Phys. Rev. A} \textbf{\bibinfo{volume}{83}},
  \bibinfo{pages}{020305(R)} (\bibinfo{year}{2011}).

\bibitem[{\citenamefont{Khodjasteh and Lidar}(2005)}]{khodj05}
\bibinfo{author}{\bibfnamefont{K.}~\bibnamefont{Khodjasteh}} \bibnamefont{and}
  \bibinfo{author}{\bibfnamefont{D.~A.} \bibnamefont{Lidar}},
  \bibinfo{journal}{Phys. Rev. Lett.} \textbf{\bibinfo{volume}{95}},
  \bibinfo{pages}{180501} (\bibinfo{year}{2005}).

\bibitem[{\citenamefont{Uhrig}(2009)}]{uhrig09b}
\bibinfo{author}{\bibfnamefont{G.~S.} \bibnamefont{Uhrig}},
  \bibinfo{journal}{Phys. Rev. Lett.} \textbf{\bibinfo{volume}{102}},
  \bibinfo{pages}{120502} (\bibinfo{year}{2009}).

\bibitem[{\citenamefont{West et~al.}(2010)\citenamefont{West, Fong, and
  Lidar}}]{west10}
\bibinfo{author}{\bibfnamefont{J.~R.} \bibnamefont{West}},
  \bibinfo{author}{\bibfnamefont{B.~H.} \bibnamefont{Fong}}, \bibnamefont{and}
  \bibinfo{author}{\bibfnamefont{D.~A.} \bibnamefont{Lidar}},
  \bibinfo{journal}{Phys. Rev. Lett.} \textbf{\bibinfo{volume}{104}},
  \bibinfo{pages}{130501} (\bibinfo{year}{2010}).

\bibitem[{\citenamefont{Skinner et~al.}(2003)\citenamefont{Skinner, Reiss, Luy,
  Khaneja, and Glaser}}]{skinn03}
\bibinfo{author}{\bibfnamefont{T.~E.} \bibnamefont{Skinner}},
  \bibinfo{author}{\bibfnamefont{T.~O.} \bibnamefont{Reiss}},
  \bibinfo{author}{\bibfnamefont{B.}~\bibnamefont{Luy}},
  \bibinfo{author}{\bibfnamefont{N.}~\bibnamefont{Khaneja}}, \bibnamefont{and}
  \bibinfo{author}{\bibfnamefont{S.~J.} \bibnamefont{Glaser}},
  \bibinfo{journal}{J. Mag. Res.} \textbf{\bibinfo{volume}{163}},
  \bibinfo{pages}{8} (\bibinfo{year}{2003}).

\bibitem[{\citenamefont{Viola and Knill}(2003)}]{viola03}
\bibinfo{author}{\bibfnamefont{L.}~\bibnamefont{Viola}} \bibnamefont{and}
  \bibinfo{author}{\bibfnamefont{E.}~\bibnamefont{Knill}},
  \bibinfo{journal}{Phys. Rev. Lett.} \textbf{\bibinfo{volume}{90}},
  \bibinfo{pages}{037901} (\bibinfo{year}{2003}).

\bibitem[{\citenamefont{Sengupta and Pryadko}(2005)}]{sengu05}
\bibinfo{author}{\bibfnamefont{P.}~\bibnamefont{Sengupta}} \bibnamefont{and}
  \bibinfo{author}{\bibfnamefont{L.~P.} \bibnamefont{Pryadko}},
  \bibinfo{journal}{Phys. Rev. Lett.} \textbf{\bibinfo{volume}{95}},
  \bibinfo{pages}{037202} (\bibinfo{year}{2005}).

\bibitem[{\citenamefont{Khodjasteh and Lidar}(2007)}]{khodj07}
\bibinfo{author}{\bibfnamefont{K.}~\bibnamefont{Khodjasteh}} \bibnamefont{and}
  \bibinfo{author}{\bibfnamefont{D.~A.} \bibnamefont{Lidar}},
  \bibinfo{journal}{Phys. Rev. A} \textbf{\bibinfo{volume}{75}},
  \bibinfo{pages}{062310} (\bibinfo{year}{2007}).

\bibitem[{\citenamefont{Pryadko and Quiroz}(2008)}]{pryad08a}
\bibinfo{author}{\bibfnamefont{L.~P.} \bibnamefont{Pryadko}} \bibnamefont{and}
  \bibinfo{author}{\bibfnamefont{G.}~\bibnamefont{Quiroz}},
  \bibinfo{journal}{Phys. Rev. A} \textbf{\bibinfo{volume}{77}},
  \bibinfo{pages}{012330} (\bibinfo{year}{2008}).

\bibitem[{\citenamefont{Du et~al.}(2009)\citenamefont{Du, Rong, Zhao, Wang,
  Yang, and Liu}}]{du09}
\bibinfo{author}{\bibfnamefont{J.}~\bibnamefont{Du}},
  \bibinfo{author}{\bibfnamefont{X.}~\bibnamefont{Rong}},
  \bibinfo{author}{\bibfnamefont{N.}~\bibnamefont{Zhao}},
  \bibinfo{author}{\bibfnamefont{Y.}~\bibnamefont{Wang}},
  \bibinfo{author}{\bibfnamefont{J.}~\bibnamefont{Yang}}, \bibnamefont{and}
  \bibinfo{author}{\bibfnamefont{R.~B.} \bibnamefont{Liu}},
  \bibinfo{journal}{Nature} \textbf{\bibinfo{volume}{461}},
  \bibinfo{pages}{1265} (\bibinfo{year}{2009}).

\bibitem[{\citenamefont{Pasini et~al.}(2009)\citenamefont{Pasini, Karbach,
  Raas, and Uhrig}}]{pasin09a}
\bibinfo{author}{\bibfnamefont{S.}~\bibnamefont{Pasini}},
  \bibinfo{author}{\bibfnamefont{P.}~\bibnamefont{Karbach}},
  \bibinfo{author}{\bibfnamefont{C.}~\bibnamefont{Raas}}, \bibnamefont{and}
  \bibinfo{author}{\bibfnamefont{G.~S.} \bibnamefont{Uhrig}},
  \bibinfo{journal}{Phys. Rev. A} \textbf{\bibinfo{volume}{80}},
  \bibinfo{pages}{022328} (\bibinfo{year}{2009}).

\bibitem[{\citenamefont{Khodjasteh and Viola}(2009)}]{khodj09a}
\bibinfo{author}{\bibfnamefont{K.}~\bibnamefont{Khodjasteh}} \bibnamefont{and}
  \bibinfo{author}{\bibfnamefont{L.}~\bibnamefont{Viola}},
  \bibinfo{journal}{Phys. Rev. Lett.} \textbf{\bibinfo{volume}{102}},
  \bibinfo{pages}{080501} (\bibinfo{year}{2009}).

\bibitem[{\citenamefont{Uhrig and Pasini}(2010)}]{uhrig10a}
\bibinfo{author}{\bibfnamefont{G.~S.} \bibnamefont{Uhrig}} \bibnamefont{and}
  \bibinfo{author}{\bibfnamefont{S.}~\bibnamefont{Pasini}},
  \bibinfo{journal}{New J. Phys.} \textbf{\bibinfo{volume}{12}},
  \bibinfo{pages}{045001} (\bibinfo{year}{2010}).

\bibitem[{enr()}]{enrg}
\bibinfo{journal}{No constraints are imposed on the total energy of all pulses
  in the sequence. We consider the bound on the amplitudes of the pulses to be
  the crucial experimental constraint. This is the main difference to Ref.\
  \protect\cite{gordo08} where the technique of optimum control by modulation
  at given total energy is developed.}

\bibitem[{\citenamefont{Schliemann et~al.}(2003)\citenamefont{Schliemann,
  Khaetskii, and Loss}}]{schli03}
\bibinfo{author}{\bibfnamefont{J.}~\bibnamefont{Schliemann}},
  \bibinfo{author}{\bibfnamefont{A.}~\bibnamefont{Khaetskii}},
  \bibnamefont{and} \bibinfo{author}{\bibfnamefont{D.}~\bibnamefont{Loss}},
  \bibinfo{journal}{J. Phys.: Condens. Matter} \textbf{\bibinfo{volume}{15}},
  \bibinfo{pages}{R1809} (\bibinfo{year}{2003}).

\bibitem[{\citenamefont{Bortz and Stolze}(2006)}]{bortz07a}
\bibinfo{author}{\bibfnamefont{M.}~\bibnamefont{Bortz}} \bibnamefont{and}
  \bibinfo{author}{\bibfnamefont{J.}~\bibnamefont{Stolze}},
  \bibinfo{journal}{J. Stat. Mech.} p. \bibinfo{pages}{P06018}
  (\bibinfo{year}{2006}).

\bibitem[{\citenamefont{Bortz and Stolze}(2007)}]{bortz07b}
\bibinfo{author}{\bibfnamefont{M.}~\bibnamefont{Bortz}} \bibnamefont{and}
  \bibinfo{author}{\bibfnamefont{J.}~\bibnamefont{Stolze}},
  \bibinfo{journal}{Phys. Rev. B} \textbf{\bibinfo{volume}{76}},
  \bibinfo{pages}{014304} (\bibinfo{year}{2007}).

\bibitem[{\citenamefont{Witzel and \mbox{Das~Sarma}}(2008)}]{witze08}
\bibinfo{author}{\bibfnamefont{W.~M.} \bibnamefont{Witzel}} \bibnamefont{and}
  \bibinfo{author}{\bibfnamefont{S.}~\bibnamefont{\mbox{Das~Sarma}}},
  \bibinfo{journal}{Phys. Rev. B} \textbf{\bibinfo{volume}{77}},
  \bibinfo{pages}{165319} (\bibinfo{year}{2008}).

\bibitem[{\citenamefont{Witzel et~al.}(2007)\citenamefont{Witzel, Carroll,
  Morello, Cywi\'nski, and \mbox{Das~Sarma}}}]{witze10}
\bibinfo{author}{\bibfnamefont{W.~M.} \bibnamefont{Witzel}},
  \bibinfo{author}{\bibfnamefont{M.~S.} \bibnamefont{Carroll}},
  \bibinfo{author}{\bibfnamefont{A.}~\bibnamefont{Morello}},
  \bibinfo{author}{\bibfnamefont{L.}~\bibnamefont{Cywi\'nski}},
  \bibnamefont{and}
  \bibinfo{author}{\bibfnamefont{S.}~\bibnamefont{\mbox{Das~Sarma}}},
  \bibinfo{journal}{Phys. Rev. Lett.} \textbf{\bibinfo{volume}{76}},
  \bibinfo{pages}{241303} (\bibinfo{year}{2007}).

\bibitem[{\citenamefont{Cummins et~al.}(2003)\citenamefont{Cummins, Llewellyn,
  and Jones}}]{cummi03}
\bibinfo{author}{\bibfnamefont{H.~K.} \bibnamefont{Cummins}},
  \bibinfo{author}{\bibfnamefont{G.}~\bibnamefont{Llewellyn}},
  \bibnamefont{and} \bibinfo{author}{\bibfnamefont{J.~A.} \bibnamefont{Jones}},
  \bibinfo{journal}{Phys. Rev. A} \textbf{\bibinfo{volume}{67}},
  \bibinfo{pages}{042308} (\bibinfo{year}{2003}).

\bibitem[{\citenamefont{Khodjasteh et~al.}(2010)\citenamefont{Khodjasteh,
  Lidar, and Viola}}]{khodj10}
\bibinfo{author}{\bibfnamefont{K.}~\bibnamefont{Khodjasteh}},
  \bibinfo{author}{\bibfnamefont{D.~A.} \bibnamefont{Lidar}}, \bibnamefont{and}
  \bibinfo{author}{\bibfnamefont{L.}~\bibnamefont{Viola}},
  \bibinfo{journal}{Phys. Rev. Lett.} \textbf{\bibinfo{volume}{104}},
  \bibinfo{pages}{090501} (\bibinfo{year}{2010}).

\bibitem[{\citenamefont{Uhrig and Lidar}(2010)}]{uhrig10b}
\bibinfo{author}{\bibfnamefont{G.~S.} \bibnamefont{Uhrig}} \bibnamefont{and}
  \bibinfo{author}{\bibfnamefont{D.~A.} \bibnamefont{Lidar}},
  \bibinfo{journal}{Phys. Rev. A} \textbf{\bibinfo{volume}{82}},
  \bibinfo{pages}{012301} (\bibinfo{year}{2010}).

\bibitem[{\citenamefont{Lidar et~al.}(2008)\citenamefont{Lidar, Zanardi, and
  Khodjasteh}}]{lidar08}
\bibinfo{author}{\bibfnamefont{D.}~\bibnamefont{Lidar}},
  \bibinfo{author}{\bibfnamefont{P.}~\bibnamefont{Zanardi}}, \bibnamefont{and}
  \bibinfo{author}{\bibfnamefont{K.}~\bibnamefont{Khodjasteh}},
  \bibinfo{journal}{Phys. Rev. A} \textbf{\bibinfo{volume}{78}},
  \bibinfo{pages}{012308} (\bibinfo{year}{2008}).

\bibitem[{\citenamefont{Uhrig}(2008)}]{uhrig08}
\bibinfo{author}{\bibfnamefont{G.~S.} \bibnamefont{Uhrig}},
  \bibinfo{journal}{New J. Phys.} \textbf{\bibinfo{volume}{10}},
  \bibinfo{pages}{083024} (\bibinfo{year}{2008}); \bibinfo{journal}{Corrigendum}
   \textbf{\bibinfo{volume}{13}},
  \bibinfo{pages}{059504} (\bibinfo{year}{2011}{\natexlab{b}}).

\bibitem[{\citenamefont{Cywi\'{n}ski et~al.}(2008)\citenamefont{Cywi\'{n}ski,
  Lutchyn, Nave, and \mbox{Das~Sarma}}}]{cywin08}
\bibinfo{author}{\bibfnamefont{L.}~\bibnamefont{Cywi\'{n}ski}},
  \bibinfo{author}{\bibfnamefont{R.~M.} \bibnamefont{Lutchyn}},
  \bibinfo{author}{\bibfnamefont{C.~P.} \bibnamefont{Nave}}, \bibnamefont{and}
  \bibinfo{author}{\bibfnamefont{S.}~\bibnamefont{\mbox{Das~Sarma}}},
  \bibinfo{journal}{Phys. Rev. B} \textbf{\bibinfo{volume}{77}},
  \bibinfo{pages}{174509} (\bibinfo{year}{2008}).

\bibitem[{\citenamefont{Biercuk and Uys}(2011)}]{bierc11a}
\bibinfo{author}{\bibfnamefont{M.~J.} \bibnamefont{Biercuk}} \bibnamefont{and}
  \bibinfo{author}{\bibfnamefont{H.}~\bibnamefont{Uys}}, p.
  \bibinfo{pages}{1012.4262} (\bibinfo{year}{2011}).

\bibitem[{\citenamefont{Levitt}(2005)}]{levit05}
\bibinfo{author}{\bibfnamefont{M.~H.} \bibnamefont{Levitt}},
  \emph{\bibinfo{title}{Spin Dynamics, Basics of Nuclear Magnetic Resonance}}
  (\bibinfo{publisher}{John Wiley \& Sons, Ltd}, \bibinfo{address}{Chichester},
  \bibinfo{year}{2005}).

\bibitem[{\citenamefont{Karbach et~al.}(2008)\citenamefont{Karbach, Pasini, and
  Uhrig}}]{karba08}
\bibinfo{author}{\bibfnamefont{P.}~\bibnamefont{Karbach}},
  \bibinfo{author}{\bibfnamefont{S.}~\bibnamefont{Pasini}}, \bibnamefont{and}
  \bibinfo{author}{\bibfnamefont{G.~S.} \bibnamefont{Uhrig}},
  \bibinfo{journal}{Phys. Rev. A} \textbf{\bibinfo{volume}{78}},
  \bibinfo{pages}{022315} (\bibinfo{year}{2008}).

\end{thebibliography}

\end{document}